\documentclass[12pt]{report}
\usepackage[utf8]{inputenc}
\usepackage{blindtext}
\usepackage[letterpaper, total={8.5in, 11in}, margin=2in]{geometry}
\usepackage{ragged2e}
\usepackage{blindtext}
\usepackage{graphicx}
\usepackage{amsmath}
\usepackage{amssymb}
\usepackage{gensymb}
\usepackage{array}
\date{}

\begin{document}
\centering{\huge  Capacity Maximization of the 6G Networks Deploying IRS\\
\vspace{24pt}
\large Mobasshir Mahbub, Raed M. Shubair}

\newpage

\RaggedRight{\textbf{\Large 1.\hspace{10pt} Introduction}}\\
\vspace{18pt}
\justifying{\noindent 
For the previous thirty years, wireless innovations and methodologies, alongside their possible applications, have undergone exponential growth and tremendous advancement. These include transmitter design as well as transmission characteristics [1-18], Thz connection and signal conditioning features [19-31], and indoor positioning strategies and difficulties [32-50].

While 5G wireless technologies [51], [52] are being implemented throughout the world, experts are beginning to anticipate 6G networks that would merge sensing, communication, computing, and control operations [53]. The 6G communication system is expected to deliver massive communication to vast numbers of interlinked gadgets with varied quality of service (QoS) prerequisites, omnipresent coverage, a high level of integrated artificial intelligence (AI), effective energy usage, and dynamic network security [54], [55]. 6G predicts optimal spectral utilization using multi-band wide-spread spectrum, along with higher radio frequencies such as Sub-6 GHz, millimeter-wave (mmWave), and terahertz (THz) bands [56] to allow extremely high data transfer rate.

The presence of enormous physical objects with wireless connectivity is motivating IoT system development [57], [58], such as Ultra-Reliable Low Latency Connectivity (URLLC), massive machine-to-machine connectivity, wireless sensor networks, etc.

As a contemporary innovation, the Internet of Things (IoT) is defined as worldwide network architecture comprised of countless linked intelligent objects that share information and coordinate activities [59]. Because IoT contains billions of linked sensors interacting over the wireless networks, the issue of power consumption of the devices appears as a major challenge and restriction for the expansion of IoT. To develop a sustainable and environmentally friendly IoT network, it is imperative to maintain equilibrium between spectrum and energy efficiency.

IRSs have attracted the attention of industry and academics as a low-cost enhancement element for 6G wireless networks that require great energy and spectrum efficiency [60]. By adjusting the reflection characteristics of a significant number of passive reflective components, an IRS may modify a wireless transmission channel to be more conducive to signal or energy transmission [61]-[63]. As a result, the IRS minimizes energy consumption with passive components while achieving full-duplex gain without requiring complex signal manipulation such as interference suppression and demodulation.

The work aimed at the enhancement of the number of supportable devices deploying IRS in a micro cell operating under a macro cell, i.e., a two-tier network.}

\vspace{18pt}

\RaggedRight{\textbf{\Large 2.\hspace{10pt} Related Literature}}\\
\vspace{18pt}
\justifying{\noindent The paper in this section includes a review of relative prior works to provide an overview of ongoing works.

Han et al. [64] investigated the capacity maximization or optimization problem by optimizing the passive beamforming and transmit covariance matrices of two interacting IRS. Papazafeiropoulos et al. [65] analyzed the ergodic capacity maximization problem of an IRS-assisted multi-input multi-output (MIMO) system. Choi et al. [66] aimed the capacity maximization of a multiple IRSs-assisted line-of-sight (LoS) MIMO network through the optimization of the phase shifter matrices of the IRS. In the context of the optimization problem, the work proposed a novel beamforming scheme. Asghar et al. [67] proposed a scheme to maximize capacity and coverage while minimizing the imbalance of the cell load among the macro and small cells. The work formulated an optimization problem in terms of transmit power, antenna tilt, and cell offset that directly influence the capacity, coverage, and cell load. Chen et al. [68] proposed a joint scheme to maximize the capacity of a network by jointly optimizing the power, polarization states, and subcarriers through an iterative technique.
}
\vspace{18pt}

\RaggedRight{\textbf{\Large 3.\hspace{10pt} Measurement Model}}\\
\vspace{18pt}
\justifying{\noindent The work considered a two-tier network formed by a macro cell base station and a tier of micro cell base stations (operating under the macro cell). The associated users or IoT devices are served by corresponding base stations. In the case of an IRS-assisted transmission model, the user devices will be served by the micro base station via an IRS.}

\vspace{12pt}
\RaggedRight{\textit{\large A.\hspace{10pt} Conventional Transmission Model}}\\
\vspace{12pt}

\justifying In a conventional/typical micro cellular transmission model the downstream or downlink received power is defined by the formula below. (Eq. 1) [69], [70],

\begin{equation}
P^{Conv}_{Rec}=P_{Tr}\frac{\lambda^2}{16\pi^2R^\alpha}
\end{equation}
where $P_{Tr}$ is the transmit power of the micro base station. The radio wavelength of the signal is $\lambda= c⁄f_c$. $c$ is the velocity or speed of light propagation ($ms^{-1}$). The carrier frequency is $f_c$ (Hz). $R= \sqrt{(x^{B}-x^{D})^2+(y^{B}-y^{D})^2+(z^{B}-z^{D})^2}$ is the distance of the transmitting station, i.e., base station and receiving IoT devices positioned at $(x^{B},y^{B},z^{B})$ and $(x^{D},y^{D},z^{D})$ coordinates. $\alpha$ is an exponent denoting the level of signal attenuation (or can be termed as the path loss exponent).

\vspace{12pt}
\RaggedRight{\textit{\large B.\hspace{10pt} IRS-Assisted Transmission Model}}\\
\vspace{12pt}

\justifying In an IRS-assisted micro cellular model, the downlink received power is obtained by the following formula (Eq. 2) [71],

\begin{equation}
P^{IRS}_{Rec} = \frac{d_x d_y M^2 N^2 \lambda^2 A^2 G_S G_T G_R   Cos\theta_T Cos\theta_R}{(D_1 D_2)^2 64\pi^3}P_{Tr}
\end{equation}
where $M$, $N$, $\theta_T$, $\theta_R$, and $A$ are denoting the number of transmitting elements (of IRS), receiving elements, transmitting angle (base station-to-IRS), receiving angle (IRS-to-receiver or device), and reflection coefficient, respectively. $d_x$ and $d_y$ indicate the length and width of each scattering element.   $d_x$ and $d_y=\lambda⁄2$ meter. $G_T$ and $G_R$ are the gains of the transmitting antenna and receiving unit or device. $G_S = \frac{d_x d_y 4\pi}{\lambda^2}$  is IRS elements’ scattering gain. $D_1= \sqrt{(x^{B}-x^{S})^2+(y^{B}-y^{S})^2+(z^{B}-z^{S})^2}$ is the spacing (distance) between the transmitting unit and the IRS situated at $(x^B,y^B,z^B)$ and $(x^S,y^S,z^S)$ coordinates, respectively.
$D_2= \sqrt{(x^{S}-x^{D})^2+(y^{S}-y^{D})^2+(z^{S}-z^{D})^2}$ indicates the space or gap between the IRS and the receiving devices situating at $(x^D,y^D,z^D)$ coordinates.

\vspace{12pt}
\RaggedRight{\textit{\large C.\hspace{10pt} User Association and Cell Capacity Analysis}}\\
\vspace{12pt}

\justifying \textbf{User Association:} The probability of user association for the conventional and IRS-assisted micro cellular transmission under a macro base station is expressed as (Eq. 3) [72],

\begin{equation}
\mathcal{A}^{Conv/IRS} = \left (1+\frac{\lambda_{Ma}}{\lambda_{Mi}}\left(\frac{P^{Ma}_{Rec}}{P^{Mi (Conv/IRS)}_{Rec}}\right)^\frac{2}{\alpha_{Ma}}\right)^{-1}
\end{equation}
where $\lambda_{Ma}$ and $\lambda_{Mi}$ indicate the densities (per $m^2$) of the macro and micro base station/s, respectively. $P_{Rec}^{Ma}$ is the interfering received power from the macro base station. $P_{Rec}^{Mi (Conv/IRS)}$ is the received power by the devices from the micro base station (conventional or IRS-assisted). $\alpha_{Ma}$ is the exponent for representing the level of signal attenuation.

\textbf{Cell Capacity:} The cell capacity namely the average number of associated devices by the micro base station (in the case of both of the models) is measured by the following equation (Eq. 4) [73],

\begin{equation}
\mathcal{N}^{Conv/IRS} = 1+ \left (1.28\lambda_u\frac{\Bar{\mathcal{A}}^{Conv/IRS}}{\lambda_{Mi}}\right)
\end{equation}
where $\lambda_u$ is the IoT device density per $m^2$. $\Bar{\mathcal{A}}^{Conv/IRS}$ is the averaged probability of user association.

\vspace{18pt}

\RaggedRight{\textbf{\Large 4.\hspace{10pt} Numerical Results and Discussions}}\\
\vspace{18pt}
\justifying
The section of the paper includes the numerical results and discussion corresponding to the computer-aided namely MATLAB-based analysis. Table 1 includes the measurement parameters and corresponding values.

\begin{table}[htbp]
\caption{Parameters and Corresponding Values}
\begin{center}
\begin{tabular}{| m{3.5cm} | m{3.5cm}|}
\hline
\textbf{\textit{Parameters}}& \textbf{\textit{Values}}\\
\hline
Area of macro cell & 1000 sq. m\\
\hline
Area of micro cell & 200 sq. m\\
\hline
Power macro BS & 50 W \\
\hline
Macro BS height & 10 m\\
\hline
Micro BS height & 5 m\\
\hline
IRS height & 6 m\\
\hline
User/device height & 1.5 m\\
\hline
Tx-rx gains & 20 dB [71] and 15 dB [74]\\
\hline
Tx-rx elements & 128, 256\\
\hline
Tx and rx angle & 45$\degree$ [71]\\
\hline
Carrier & 30, 55, 90, 120 GHz\\
\hline
IRS's refl. coefficient & 0.9 [71]\\
\hline
Density of miBSs & $1000⁄\pi(100)^2$  per $m^2$\\ 
\hline
Density of maBSs & $(1000⁄\pi(100)^2)/5$  per $m^2$\\
\hline
Density of devices & $(1000⁄\pi(100)^2)\times500$  per $m^2$ (max)\\
\hline
Micro BS attenuation exponent & 2.5\\
\hline
Macro BS attenuation exponent & 4.5\\
\hline
\end{tabular}
\label{tab1}
\end{center}
\end{table}

Fig. 1 illustrates the probability of user association in the case of a conventional network in terms of transmitter-receiver separation for selected mmWave carriers.

\begin{figure}[htbp]
\centerline{\includegraphics[height=6.0cm, width=8.0cm]{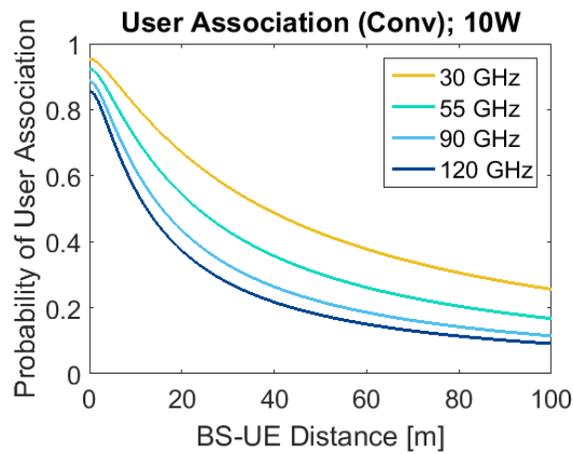}}
\caption{Probability of user association for the conventional network.}
\end{figure}

Fig. 2 shows the analysis of the probability of user association in the case of an IRS-assisted network in terms of transmitter-receiver separation distance.

\begin{figure}[htbp]
\centerline{\includegraphics[height=6.0cm, width=8.0cm]{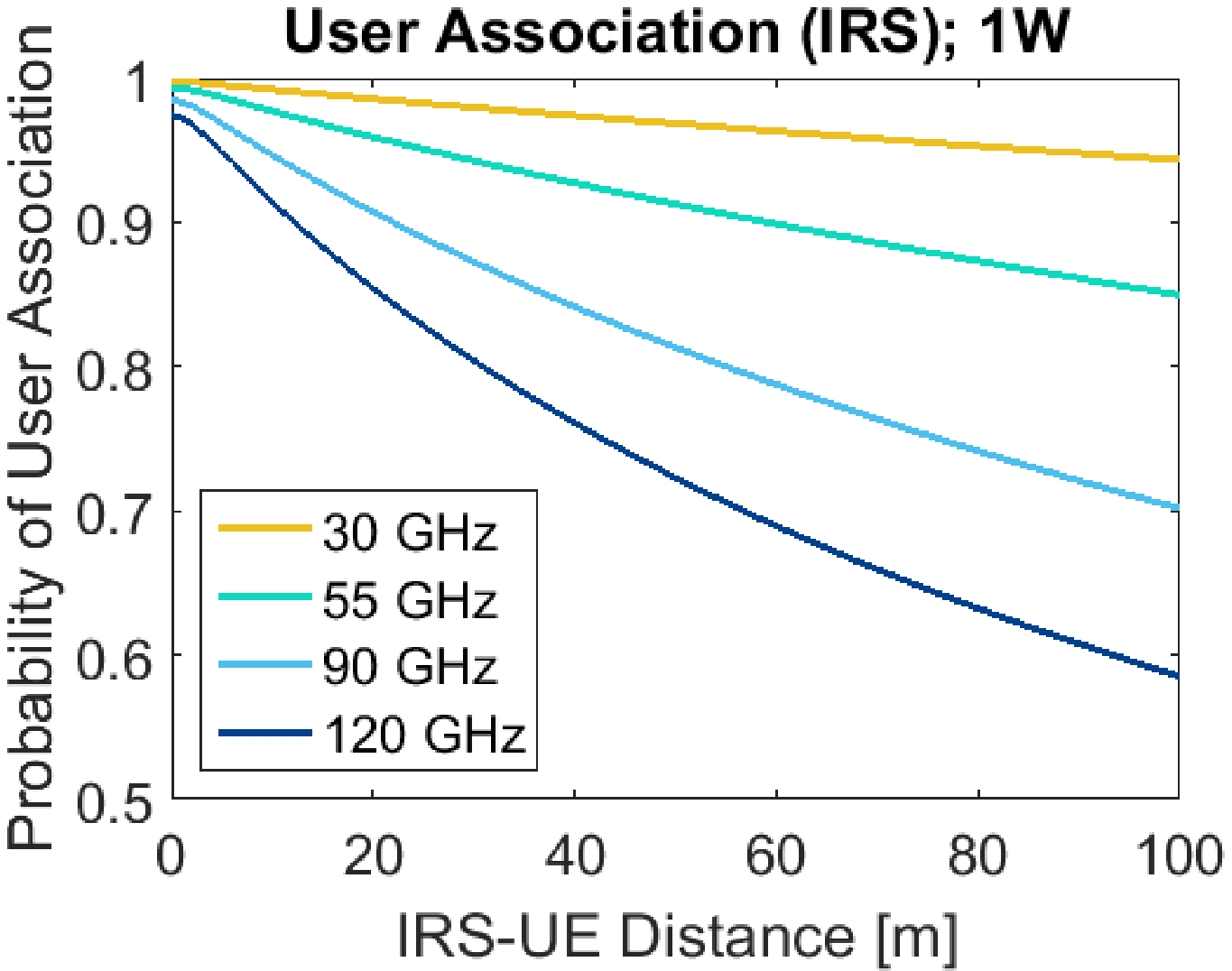}}
\caption{Probability of user association for IRS-assisted network (256 transmitter-receiver elements).}
\end{figure}

Fig. 3 visualizes the number of IoT devices that can be served the micro cell base station in the case of a conventional network in terms of user device density (per $m^2$).

\begin{figure}[htbp]
\centerline{\includegraphics[height=6.0cm, width=8.0cm]{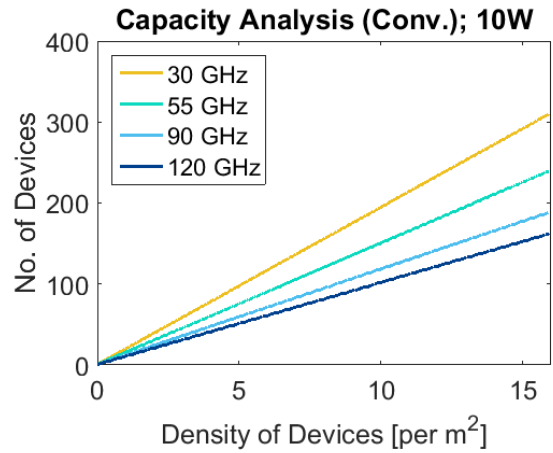}}
\caption{Number of supportable devices in a conventional network.}
\end{figure}

Fig. 4 (a)-(d) shows the number of users or IoT devices that can be served the IRS-assisted micro cell base station in terms of the density of devices (per $m^2$).

\begin{figure}[htbp]
\centerline{\includegraphics[height=6.0cm, width=8.0cm]{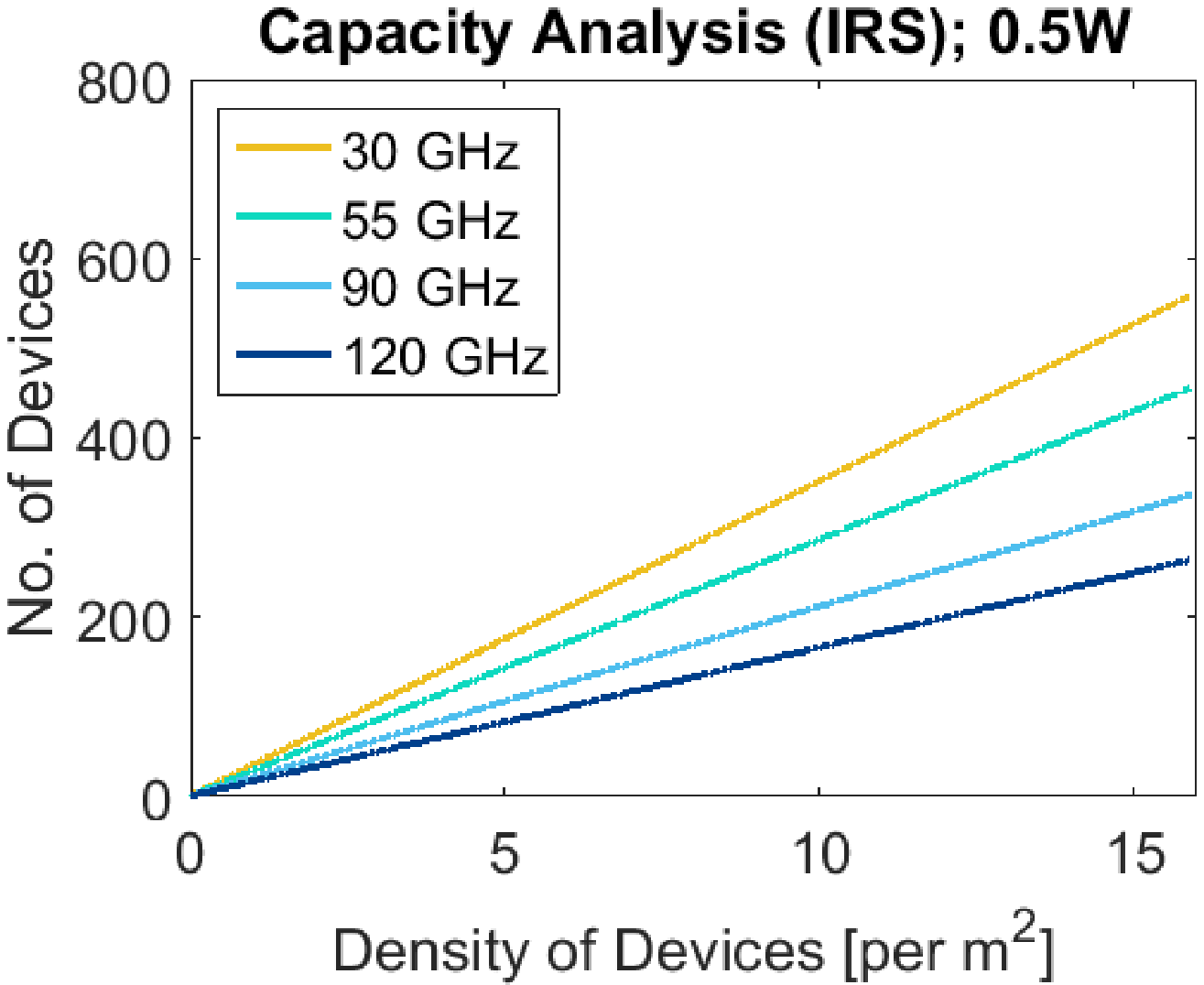}}
\vspace{3pt}
\centerline{\footnotesize{(a)}}
\label{fig}
\end{figure}

\begin{figure}[htbp]
\centerline{\includegraphics[height=6.0cm, width=8.0cm]{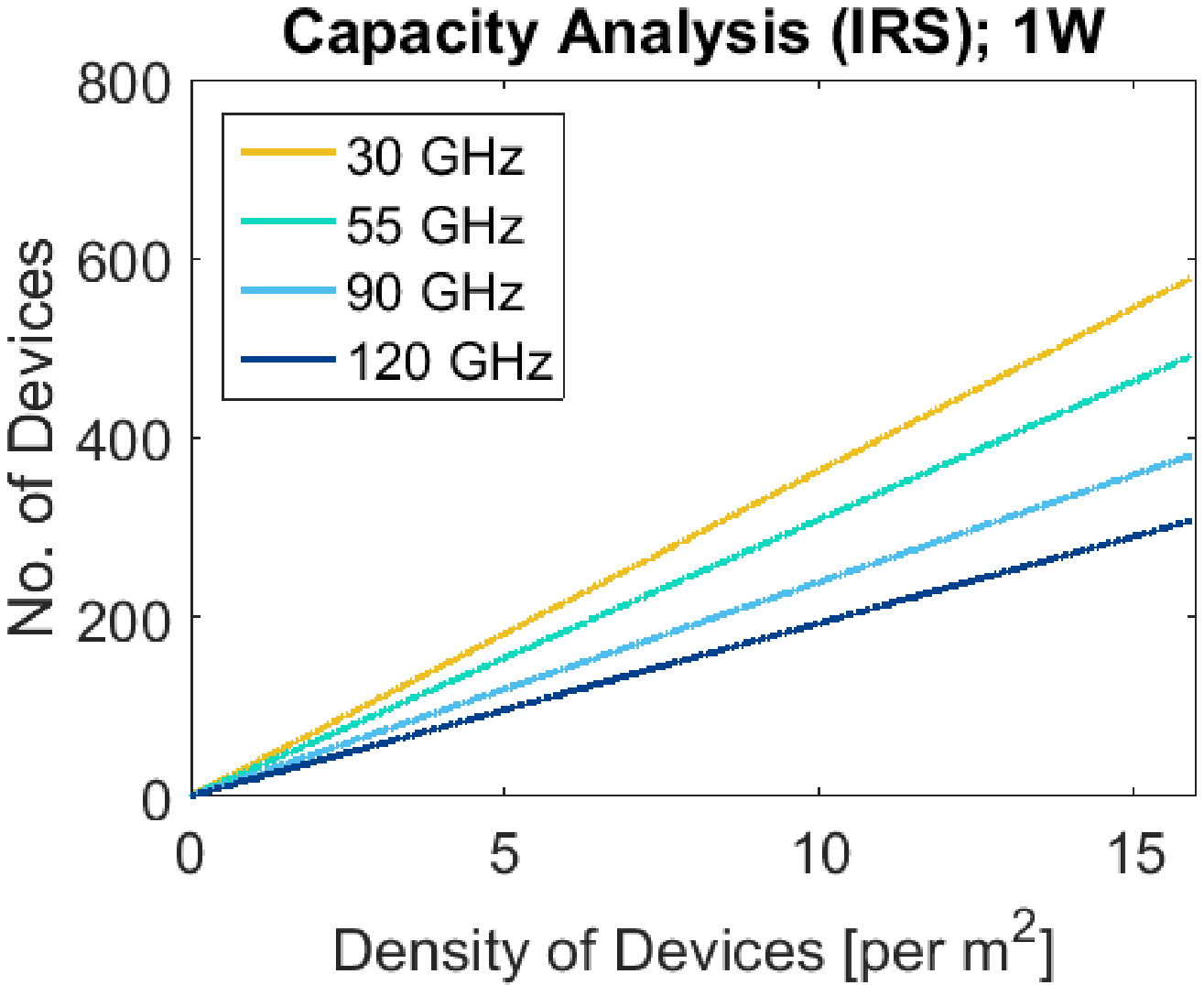}}
\vspace{3pt}
\centerline{\footnotesize{(b)}}
\label{fig}
\end{figure}

\begin{figure}[htbp]
\centerline{\includegraphics[height=6.0cm, width=8.0cm]{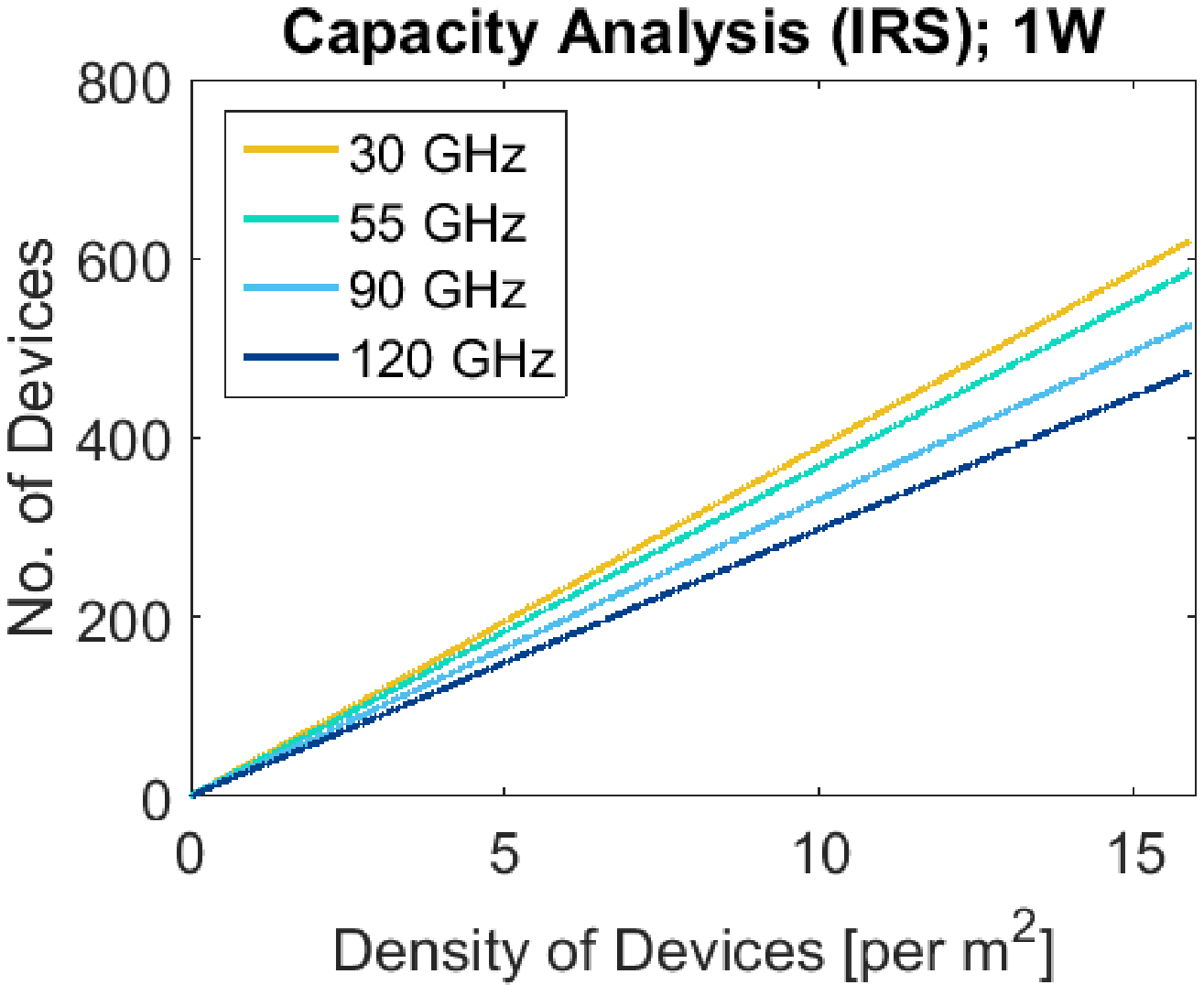}}
\vspace{3pt}
\centerline{\footnotesize{(c)}}
\label{fig}
\end{figure}

\begin{figure}[htbp]
\centerline{\includegraphics[height=6.0cm, width=8.0cm]{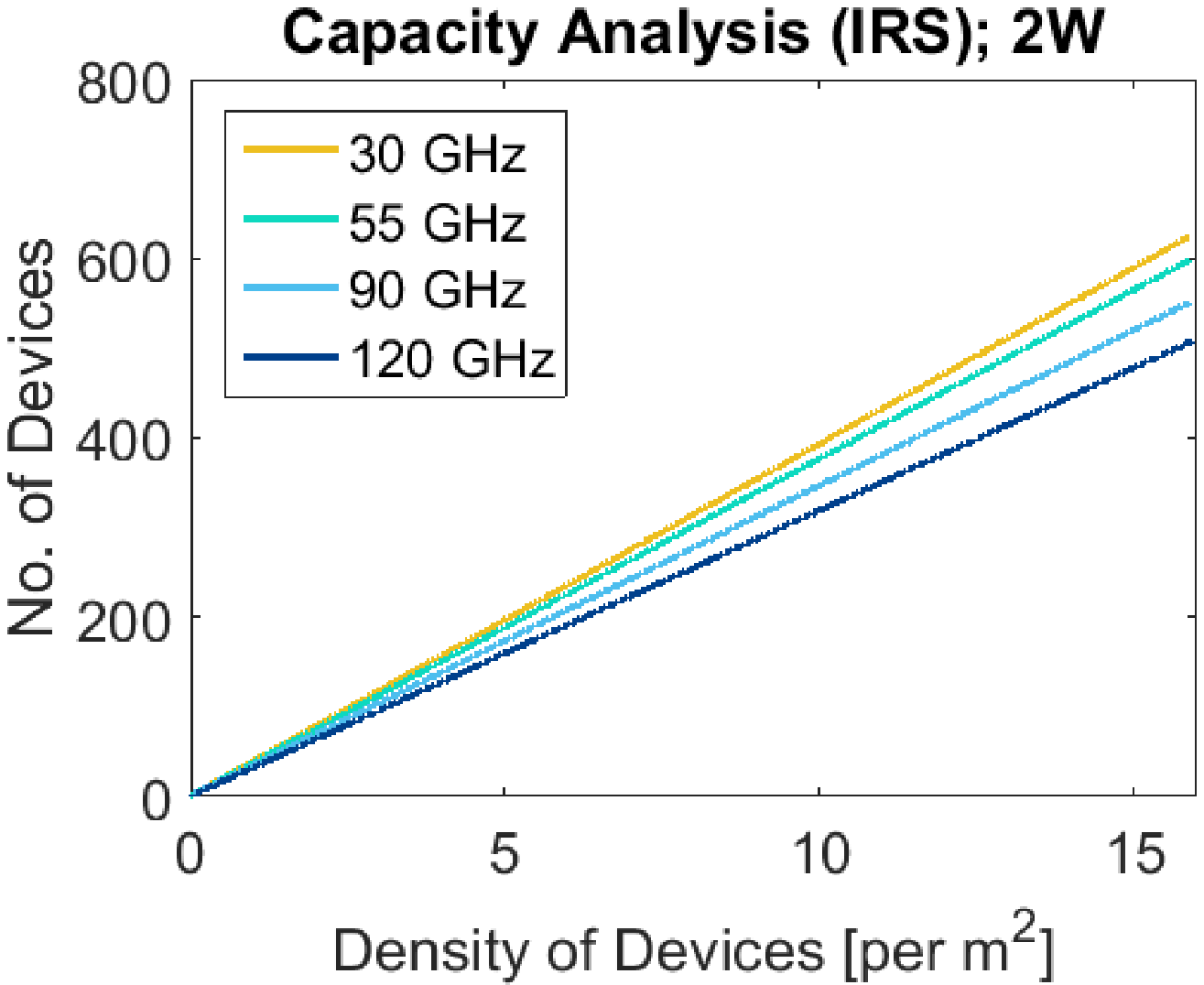}}
\vspace{3pt}
\centerline{\footnotesize{(d)}}
\caption{Number of supportable devices in an IRS-assisted network (a) for 0.5 W transmits power and 128 transmitter-receiver elements, (b) for 1 W and 128 elements, (c) for 1 W and 256 elements, (d) for 2 W and 256 elements.}
\label{fig}
\end{figure}

The comparative study between Fig. 1 and 2 defines that the deployment of IRS in a micro cell improves the user association performance significantly.

In the case of a conventional network (Fig. 3), 30, 55, 90, and 120 GHz mmWave carriers can support up to 310, 239, 188, and 162 users (IoT devices) in terms of the selected density (per $m^2$).

The observation of Fig. 4 (a) defines that, 30, 55, 90, and 120 GHz mmWave carriers can support up to 559, 455, 337, and 265 devices, respectively (in terms of the chosen density of devices for analysis).

From Fig. 4 (b) it is realizable that, 30, 55, 90, and 120 GHz mmWave carriers can support up to 578, 491, 381, and 308 devices, respectively.

Studying the findings from Fig. 4 (c) it is comprehensible that, 30, 55, 90, and 120 GHz mmWave carriers can support up to 621, 586, 525, and 473 devices, respectively.

Fig. 4 (d) states that 30, 55, 90, and 120 GHz mmWave carriers can support up to 626, 599, 552, and 506 devices, respectively.

The number of supportable devices got increased with the increase of the transmit power of the micro base station (from the comparison of Fig. 4(a) and (b)).

However, the comparison of Fig. 4 (b) and (c) states that the performance of the network improved drastically when the number of transmitter-receiver elements of IRS is increased. Especially the performance of the higher-level carriers enhanced significantly. 43, 95, 144, and 165 more devices can be supported (Fig. 4(c)) by the micro base station increasing the number of transmitter-receiver elements of IRS for 30, 55, 90, and 120 GHz mmWave carriers, respectively with the same transmit power (of Fig. 4(b)).

From the comparison of Fig. 3 and 4, it is realizable that, through the deployment of IRS an increased number of devices can be supported by the micro base station with a significant reduction of energy consumption. With 80-95\% reduced transmit power of the micro base station the performance of the network got improved when an IRS is deployed.

\vspace{18pt}

\RaggedRight{\textbf{\Large 5.\hspace{10pt} Conclusion}}\\
\vspace{18pt}
\justifying \noindent {The research aimed to improve the cell capacity, i.e., the enhancement of the number of supportable users deploying IRS in a micro cell of a two-tier network. The paper included a review of prior works to provide insight into ongoing research on the relevant topic. Afterward, the work formulated a measurement model incorporating the equations to measure the capacity of the micro base station for both network models. The work derived that the capacity performance can significantly be improved by the deployment of IRS and efficient orchestration of transmit power of the micro base station and transmitter-receiver elements of IRS. Another notable finding of the research is that the implementation of IRS assures significant energy efficiency reducing the power consumption of the micro base station.}
\vspace{18pt}

\RaggedRight{\textbf{\Large References}}\\
\vspace{12pt}

\justifying{
1.	Yazan Al-Alem, Ahmed A. Kishk, and Raed Shubair. "Wireless chip to chip communication link budget enhancement using hard/soft surfaces." In 2018 IEEE Global Conference on Signal and Information Processing (GlobalSIP), pp. 1013-1014. IEEE, 2018.

2.	Melissa Eugenia Diago-Mosquera, Alejandro Aragón-Zavala, Fidel Alejandro Rodríguez-Corbo, Mikel Celaya-Echarri, Raed M. Shubair, and Leyre Azpilicueta. "Tuning Selection Impact on Kriging-Aided In-Building Path Loss Modeling." IEEE Antennas and Wireless Propagation Letters 21, no. 1 (2021): 84-88.

3.	Asimina Kiourti, and Raed M. Shubair. "Implantable and ingestible sensors for wireless physiological monitoring: a review." In 2017 IEEE International Symposium on Antennas and Propagation \& USNC/URSI National Radio Science Meeting, pp. 1677-1678. IEEE, 2017.

4.	Yazan Al-Alem, Raed M. Shubair, and Ahmed Kishk. "Efficient on-chip antenna design based on symmetrical layers for multipath interference cancellation." In 2016 16th Mediterranean Microwave Symposium (MMS), pp. 1-3. IEEE, 2016.

5.	Yazan Al-Alem, Ahmed A. Kishk, and Raed M. Shubair. "Enhanced wireless interchip communication performance using symmetrical layers and soft/hard surface concepts." IEEE Transactions on Microwave Theory and Techniques 68, no. 1 (2019): 39-50.

6.	Raed M. Shubair, Amer Salah, and Alaa K. Abbas. "Novel implantable miniaturized circular microstrip antenna for biomedical telemetry." In 2015 IEEE International Symposium on Antennas and Propagation \& USNC/URSI National Radio Science Meeting, pp. 947-948. IEEE, 2015.

7.	Hari Shankar Singh, SachinKalraiya, Manoj Kumar Meshram, and Raed M. Shubair. "Metamaterial inspired CPW‐fed compact antenna for ultrawide band applications." International Journal of RF and Microwave Computer Aided Engineering 29, no. 8 (2019): e21768.

8.	Muhammad S. Khan, Adnan Iftikhar, Raed M. Shubair, Antonio D. Capobianco, Benjamin D. Braaten, and Dimitris E. Anagnostou. "A four element, planar, compact UWB MIMO antenna with WLAN band rejection capabilities." Microwave and Optical Technology Letters 62, no. 10 (2020): 3124-3131.

9.	Muhammad Saeed Khan, Adnan Iftikhar, Raed M. Shubair, Antonio-Daniele Capobianco, Sajid Mehmood Asif, Benjamin D. Braaten, and Dimitris E. Anagnostou. "Ultra-compact reconfigurable band reject UWB MIMO antenna with four radiators." Electronics 9, no. 4 (2020): 584.

10.	Mohammed S Al-Basheir, Raed M Shubai, and Sami M. Sharif. "Measurements and analysis for signal attenuation through date palm trees at 2.1 GHz frequency." (2006).

11.	M. S. Khan, F. Rigobello, Bilal Ijaz, E. Autizi, A. D. Capobianco, R. Shubair, and S. A. Khan. "Compact 3‐D eight elements UWB‐MIMO array." Microwave and Optical Technology Letters 60, no. 8 (2018): 1967-1971.

12.	Muhammad S. Khan, Syed A. Naqvi, Adnan Iftikhar, Sajid M. Asif, Adnan Fida, and Raed M. Shubair. "A WLAN band notched compact four element UWB MIMO antenna." International Journal of RF and Microwave ComputerAided Engineering 30, no. 9 (2020): e22282.

13.	Raed M. Shubair, Amna M. AlShamsi, Kinda Khalaf, and Asimina Kiourti. "Novel miniature wearable microstrip antennas for ISM-band biomedical telemetry." In 2015 Loughborough Antennas \& Propagation Conference (LAPC), pp. 1-4. IEEE, 2015.

14.	Amjad Omar, and Raed Shubair. "UWB coplanar waveguide-fed-coplanar strips spiral antenna." In 2016 10th European Conference on Antennas and Propagation (EuCAP), pp. 1-2. IEEE, 2016.

15.	Muhammad Saeed Khan, Adnan Iftikhar, Antonio‐Daniele Capobianco, Raed M. Shubair, and Bilal Ijaz. "Pattern and frequency reconfiguration of patch antenna using PIN diodes." Microwave and Optical Technology Letters 59, no. 9 (2017): 2180-2185.

16.	M. Saeeed Khan, Adnan Iftikhar, Sajid M. Asif, Antonio Daniele Capobianco, and Benjamin D. Braaten. "A compact four elements UWB MIMO antenna with on‐demand WLAN rejection." Microwave and Optical Technology Letters 58, no. 2 (2016): 270-276.

17.	Ahmed A. Ibrahim,  JanMachac, and Raed M. Shubair. "Compact UWB MIMO antenna with pattern diversity and band rejection characteristics." Microwave and Optical Technology Letters 59, no. 6 (2017): 1460-1464.

18.	Abdul Karim Gizzini, Marwa Chafii, Shahab Ehsanfar, and Raed M. Shubair. "Temporal Averaging LSTM-based Channel Estimation Scheme for IEEE 802.11 p Standard." arXiv preprint arXiv:2106.04829 (2021).

19.	Hadeel Elayan, Raed M. Shubair, Josep M. Jornet, Asimina Kiourti, and Raj Mittra. "Graphene-Based Spiral Nanoantenna for Intrabody Communication at Terahertz." In 2018 IEEE International Symposium on Antennas and Propagation \& USNC/URSI National Radio Science Meeting, pp. 799-800. IEEE, 2018.

20.	Abdul Karim Gizzini, Marwa Chafii, Ahmad Nimr, Raed M. Shubair, and Gerhard Fettweis. "Cnn aided weighted interpolation for channel estimation in vehicular communications." IEEE Transactions on Vehicular Technology 70, no. 12 (2021): 12796-12811.

21.	Hadeel Elayan, Cesare Stefanini, Raed M. Shubair, and Josep M. Jornet. "Stochastic noise model for intra-body terahertz nanoscale communication." In Proceedings of the 5th ACM International Conference on Nanoscale Computing and Communication, pp. 1-6. 2018.

22.	Mayar Lotfy, Raed M. Shubair, Nassir Navab, and Shadi Albarqouni. "Investigation of focal loss in deep learning models for femur fractures classification." In 2019 International Conference on Electrical and Computing Technologies and Applications (ICECTA), pp. 1-4. IEEE, 2019.

23.	Hadeel Elayan, Raed M. Shubair, and Josep M. Jornet. "Characterising THz propagation and intrabody thermal absorption in iWNSNs." IET Microwaves, Antennas \& Propagation 12, no. 4 (2018): 525-532.

24.	Hadeel Elayan, and Raed M. Shubair. "On channel characterization in human body communication for medical monitoring systems." In 2016 17th International Symposium on Antenna Technology and Applied Electromagnetics (ANTEM), pp. 1-2. IEEE, 2016.

25.	Hadeel Elayan, Raed M. Shubair, Akram Alomainy, and Ke Yang. "In-vivo terahertz em channel characterization for nano-communications in wbans." In 2016 IEEE International Symposium on Antennas and Propagation (APSURSI), pp. 979-980. IEEE, 2016.

26.	Hadeel Elayan, Raed M. Shubair, and Asimina Kiourti. "On graphene-based THz plasmonic nano-antennas." In 2016 16th mediterranean microwave symposium (MMS), pp. 1-3. IEEE, 2016.

27.	Samar Elmeadawy and Raed M. Shubair. "6G wireless communications: Future technologies and research challenges." In 2019 international conference on electrical and computing technologies and applications (ICECTA), pp. 1-5. IEEE, 2019.

28.	Hadeel Elayan, Pedram Johari, Raed M. Shubair, and Josep Miquel Jornet. "Photothermal modeling and analysis of intrabody terahertz nanoscale communication." IEEE transactions on nanobioscience 16, no. 8 (2017): 755-763.

29.	Hadeel Elayan, Raed M. Shubair, and Asimina Kiourti. "Wireless sensors for medical applications: Current status and future challenges." In 2017 11th European Conference on Antennas and Propagation (EUCAP), pp. 2478-2482. IEEE, 2017.

30.	Raed M. Shubair and Hadeel Elayan. "In vivo wireless body communications: State-of-the-art and future directions." In 2015 Loughborough Antennas \& Propagation Conference (LAPC), pp. 1-5. IEEE, 2015.

31.	Mohamed I. AlHajri, Raed M. Shubair, and Marwa Chafii. "Indoor Localization Under Limited Measurements: A Cross-Environment Joint Semi-Supervised and Transfer Learning Approach." In 2021 IEEE 22nd International Workshop on Signal Processing Advances in Wireless Communications (SPAWC), pp. 266-270. IEEE, 2021.

32.	Wafa Njima, Marwa Chafii, and Raed M. Shubair. "Gan based data augmentation for indoor localization using labeled and unlabeled data." In 2021 International Balkan Conference on Communications and Networking (BalkanCom), pp. 36-39. IEEE, 2021.

33.	M. I. AlHajri, N. T. Ali, and R. M. Shubair. "2.4 ghz indoor channel measurements data set." UCI Machine Learning Repository (2018).

34.	Mohamed Ibrahim Alhajri, N. T. Ali, and R. M. Shubair. "2.4 ghz indoor channel measurements." IEEE Dataport (2018).

35.	R. M. Shubair, and A. Merri. "A convergence study of adaptive beamforming algorithms used in smart antenna systems." In 11th International Symposium on Antenna Technology and Applied Electromagnetics [ANTEM 2005], pp. 1-5. IEEE, 2005.

36.	Wafa Njima, Marwa Chafii, Arsenia Chorti, Raed M. Shubair, and H. Vincent Poor. "Indoor localization using data augmentation via selective generative adversarial networks." IEEE Access 9 (2021): 98337-98347.

37.	Raed M. Shubair, Abdulrahman S. Goian, Mohamed I. AlHajri, and Ahmed R. Kulaib. "A new technique for UCA-based DOA estimation of coherent signals." In 2016 16th Mediterranean Microwave Symposium (MMS), pp. 1-3. IEEE, 2016.

38.	Satish R. Jondhale, Raed Shubair, Rekha P. Labade, Jaime Lloret, and Pramod R. Gunjal. "Application of supervised learning approach for target localization in wireless sensor network." In Handbook of Wireless Sensor Networks: Issues and Challenges in Current Scenario's, pp. 493-519. Springer, Cham, 2020.

39.	Mohamed I. AlHajri, Nazar T. Ali, and Raed M. Shubair. "A machine learning approach for the classification of indoor environments using RF signatures." In 2018 IEEE Global Conference on Signal and Information Processing (GlobalSIP), pp. 1060-1062. IEEE, 2018.

40.	Zhenghua Chen, Mohamed I. AlHajri, Min Wu, Nazar T. Ali, and Raed M. Shubair. "A novel real-time deep learning approach for indoor localization based on RF environment identification." IEEE Sensors Letters 4, no. 6 (2020): 1-4.

41.	Raed M. Shubair, and Ali Hakam. "Adaptive beamforming using variable step-size LMS algorithm with novel ULA array configuration." In 2013 15th IEEE International Conference on Communication Technology, pp. 650-654. IEEE, 2013.

42.	Mohamed AlHajri, Abdulrahman Goian, Muna Darweesh, Rashid AlMemari, Raed Shubair, Luis Weruaga, and Ahmed AlTunaiji. "Accurate and robust localization techniques for wireless sensor networks." arXiv preprint arXiv:1806.05765 (2018).

43.	E. M. Al-Ardi, Raed M. Shubair, and M. E. Al-Mualla. "Performance evaluation of direction finding algorithms for adapative antenna arrays." In 10th IEEE International Conference on Electronics, Circuits and Systems, 2003. ICECS 2003. Proceedings of the 2003, vol. 2, pp. 735-738. IEEE, 2003.

44.	R. M. Shubair and W. Jessmi. "Performance analysis of SMI adaptive beamforming arrays for smart antenna systems." In 2005 IEEE Antennas and Propagation Society International Symposium, vol. 1, pp. 311-314. IEEE, 2005.

45.	R. M. Shubair and Y. L. Chow. "A closed-form solution of vertical dipole antennas above a dielectric half-space." IEEE transactions on antennas and propagation 41, no. 12 (1993): 1737-1741.

46.	R. M. Shubair and A. Al-Merri. "Robust algorithms for direction finding and adaptive beamforming: performance and optimization." In The 2004 47th Midwest Symposium on Circuits and Systems, 2004. MWSCAS'04., vol. 2, pp. II-II. IEEE, 2004.

47.	Ebrahim M. Al-Ardi, Raed M. Shubair, and Mohammed E. Al-Mualla. "Direction of arrival estimation in a multipath environment: An overview and a new contribution." Applied Computational Electromagnetics Society Journal 21, no. 3 (2006): 226.

48.	Fahad Belhoul, Raed M. Shubair, and Mohammed E. Al-Mualla. "Modelling and performance analysis of DOA estimation in adaptive signal processing arrays." In ICECS, pp. 340-343. 2003.

49.	Mohamed I. AlHajri, Nazar T. Ali, and Raed M. Shubair. "Indoor localization for IoT using adaptive feature selection: A cascaded machine learning approach." IEEE Antennas and Wireless Propagation Letters 18, no. 11 (2019): 2306-2310.

50.	Mohamed I. AlHajri, Nazar T. Ali, and Raed M. Shubair. "Classification of indoor environments for IoT applications: A machine learning approach." IEEE Antennas and Wireless Propagation Letters 17, no. 12 (2018): 2164-2168.

51. B. S. Khan, S. Jangsher, A. Ahmed and A. Al-Dweik, "URLLC and eMBB in 5G Industrial IoT: A Survey," in IEEE Open Journal of the Communications Society, vol. 3, pp. 1134-1163, 2022.

52. H. Fourati, R. Maaloul and L. Chaari, “A survey of 5G network systems: challenges and machine learning approaches,” International Journal of Machine Learning and Cybernetics, vol. 12, pp. 385–431, August 2020.

53. S. Elmeadawy and R. M. Shubair, "6G Wireless Communications: Future Technologies and Research Challenges," 2019 International Conference on Electrical and Computing Technologies and Applications (ICECTA), 2019, pp. 1-5.

54. C. D. Alwis et al., "Survey on 6G Frontiers: Trends, Applications, Requirements, Technologies and Future Research," in IEEE Open Journal of the Communications Society, vol. 2, pp. 836-886, 2021.

55. W. Jiang, B. Han, M. A. Habibi and H. D. Schotten, "The Road Towards 6G: A Comprehensive Survey," in IEEE Open Journal of the Communications Society, vol. 2, pp. 334-366, 2021.

56. H. Elayan, O. Amin, B. Shihada, R. M. Shubair and M. -S. Alouini, "Terahertz Band: The Last Piece of RF Spectrum Puzzle for Communication Systems," in IEEE Open Journal of the Communications Society, vol. 1, pp. 1-32, 2020.

57. A. H. Mohd Aman, E. Yadegaridehkordi, Z. S. Attarbashi, R. Hassan and Y. -J. Park, "A Survey on Trend and Classification of Internet of Things Reviews," in IEEE Access, vol. 8, pp. 111763-111782, 2020.

58. F. Guo, F. R. Yu, H. Zhang, X. Li, H. Ji and V. C. M. Leung, "Enabling Massive IoT Toward 6G: A Comprehensive Survey," in IEEE Internet of Things Journal, vol. 8, no. 15, pp. 11891-11915, 1 Aug.1, 2021.

59. C. Paniagua and J. Delsing, "Industrial Frameworks for Internet of Things: A Survey," in IEEE Systems Journal, vol. 15, no. 1, pp. 1149-1159, March 2021.

60. M. A. S. Sejan et al., “Machine Learning for Intelligent-Reflecting-Surface-Based Wireless Communication towards 6G: A Review,” Sensors, vol. 22, no. 14, July 2022.

61. R. Liu, Q. Wu, M. Di Renzo and Y. Yuan, "A Path to Smart Radio Environments: An Industrial Viewpoint on Reconfigurable Intelligent Surfaces," in IEEE Wireless Communications, vol. 29, no. 1, pp. 202-208, February 2022.

62. Q. Wu and R. Zhang, "Intelligent Reflecting Surface Enhanced Wireless Network via Joint Active and Passive Beamforming," in IEEE Transactions on Wireless Communications, vol. 18, no. 11, pp. 5394-5409, Nov. 2019.

63. B. Zheng, C. You, W. Mei and R. Zhang, "A Survey on Channel Estimation and Practical Passive Beamforming Design for Intelligent Reflecting Surface Aided Wireless Communications," in IEEE Communications Surveys \& Tutorials, vol. 24, no. 2, pp. 1035-1071, Secondquarter 2022.

64. Y. Han, S. Zhang, L. Duan and R. Zhang, "Double-IRS Aided MIMO Communication Under LoS Channels: Capacity Maximization and Scaling," in IEEE Transactions on Communications, vol. 70, no. 4, pp. 2820-2837, April 2022.

65. A. Papazafeiropoulos, "Ergodic Capacity of IRS-Assisted MIMO Systems With Correlation and Practical Phase-Shift Modeling," in IEEE Wireless Communications Letters, vol. 11, no. 2, pp. 421-425, Feb. 2022.

66. J. Choi, G. Kwon and H. Park, "Multiple Intelligent Reflecting Surfaces for Capacity Maximization in LOS MIMO Systems," in IEEE Wireless Communications Letters, vol. 10, no. 8, pp. 1727-1731, Aug. 2021.

67. A. Asghar, H. Farooq and A. Imran, "Concurrent Optimization of Coverage, Capacity, and Load Balance in HetNets Through Soft and Hard Cell Association Parameters," in IEEE Transactions on Vehicular Technology, vol. 67, no. 9, pp. 8781-8795, Sept. 2018.

68. S. Chen, Z. Zeng and C. Guo, "Exploiting Polarization for System Capacity Maximization in Ultra-Dense Small Cell Networks," in IEEE Access, vol. 5, pp. 17059-17069, 2017.

69. T. Mir, L. Dai, Y. Yang, W. Shen and B. Wang, "Optimal FemtoCell Density for Maximizing Throughput in 5G Heterogeneous Networks under Outage Constraints," 2017 IEEE 86th Vehicular Technology Conference (VTC-Fall), Toronto, ON, Canada, 2017, pp. 1-5.

70. N. Hassan and X. Fernando, "Interference Mitigation and Dynamic User Association for Load Balancing in Heterogeneous Networks," in IEEE Transactions on Vehicular Technology, vol. 68, no. 8, pp. 7578-7592, Aug. 2019.

71. W. Tang et al., "Wireless Communications With Reconfigurable Intelligent Surface: Path Loss Modeling and Experimental Measurement," in IEEE Transactions on Wireless Communications, vol. 20, no. 1, pp. 421-439.

72. M. O. Al-Kadri, Y. Deng, A. Aijaz and A. Nallanathan, "Full-Duplex Small Cells for Next Generation Heterogeneous Cellular Networks: A Case Study of Outage and Rate Coverage Analysis," in IEEE Access, vol. 5, pp. 8025-8038, 2017.

73. M. M. Fadoul, “Rate and Coverage Analysis in Multi-tier Heterogeneous Network Using Stochastic Geometry Approach,” in Ad Hoc Networks, vol.  98. 102038, Mar. 2020.

74.  F.-P. Lai, S.-Y. Mi, Y.-S. Chen, "Design and Integration of Millimeter-Wave 5G and WLAN Antennas in Perfect Full-Screen Display Smartphones" Electronics, vol. 11, no. 6, March 2022.

}

\end{document}